\documentclass[aps,prl,showpacs,twocolumn,amsmath,amsfonts,amssymb,superscriptaddress,tightenlines]{revtex4-1}

\usepackage[colorlinks,citecolor=blue,linkcolor=cyan]{hyperref}
\usepackage{grffile}
\usepackage[caption=false]{subfig}
\usepackage{graphicx}
\usepackage{mathtools}
\usepackage{scalerel}

\usepackage{todonotes}

\renewcommand{\paragraph}[1]{{\par\it #1---}\ignorespaces}
\usepackage{xcolor}
\usepackage{paralist}
\newcommand{\dg}{^{\dagger}}

\renewcommand{\vec}[1]{{\boldsymbol{#1}}}
\newcommand{\op}{\hat}

\newcommand{\wt}{\widetilde}

\newcommand{\pr}{^{\prime}}

\newcommand{\smat}[1]{\left(
    \begin{smallmatrix}
      #1
    \end{smallmatrix}
  \right)}

\begin{document}
\date{\today}

\title{Non-dispersing wave packets in lattice Floquet systems}

\author{Zhoushen Huang}%
\email{zsh@anl.gov}
\affiliation{Materials Science Division, Argonne National Laboratory}%
\author{Aashish Clerk}%
\email{aaclerk@uchicago.edu}%
\affiliation{Pritzker School of Molecular Engineering, University of Chicago}%
\author{Ivar Martin}%
\email{ivar@anl.gov}%
\affiliation{Materials Science Division, Argonne National Laboratory}%

\begin{abstract}
  We show that in a one-dimensional translationally invariant tight
  binding chain, non-dispersing wave packets can in general be
  realized as Floquet eigenstates---or linear combinations
  thereof---using a spatially inhomogeneous drive, which can be as
  simple as modulation on a single site.  The recurrence time of these
  wave packets (their ``round trip'' time) locks in at rational ratios
  $sT/r$ of the driving period $T$, where $s,r$ are co-prime
  integers. Wave packets of different $s/r$ can co-exist under the
  same drive, yet travel at different speeds. They retain their
  spatial compactness either infinitely ($s/r=1$) or over long time
  ($s/r \neq 1$). Discrete time translation symmetry is manifestly
  broken for $s \neq 1$, reminiscent of integer and fractional Floquet
  time crystals.  We further demonstrate how to reverse-engineer a
  drive protocol to reproduce a target Floquet micromotion, such as
  the free propagation of a wave packet, as if coming from a strictly
  linear energy spectrum. The variety of control schemes open up a new
  avenue for Floquet engineering in quantum information sciences.
\end{abstract}

\maketitle

\paragraph{Introduction}
It is well known that under a time-independent Hamiltonian, quantum
wave packets typically spread out due to the presence of dispersion
\cite{messiah64}.  Since the birth of quantum mechanics, the stark
contrast between the elusiveness of localized quantum entities and the
stability of their classical counterparts has motivated physicists to
explore ways to understand and even this disparity
\cite{schrodinger26, nauenberg98, buchleitner02, berry78}. Besides
conceptual interest, such dynamically stable compact entities would
hold technological utility in quantum information processing and
computing platforms, since most control technologies are local in
nature.

There are two known strategies for stabilization of non-dispersing
wave packets. The first one relies on introducing some form of
nonlinearity into the wave equation. For example, non-linear
Schr\"odinger or Gross-Pitaevskii equations are known to host soliton
solutions \cite{knight92,kivshar89}. The second one utilizes
time-periodic Hamiltonian in the conventional linear Schr\"odinger
equation, where recurring wave packets have been achieved with the
help of Floquet engineering \cite{buchleitner95, holthaus95,
  buchleitner02, maeda04, kalinski05, vela05, sacha15, goussev18}.
There, periodic driving serves to periodically reshape the wavepacket,
curbing the irreversible dispersive spread characteristic of undriven
systems.  For example, this approach was applied in the context of
microwave-driven Rydberg atoms \cite{buchleitner02, maeda04,
  dunning09} to show that some Floquet eigenstates correspond to wave
packets following classical Kepler orbits.  While the shape of these
wave packets is time-dependent, they refocus almost perfectly after
integer number of drive periods. Intra-period, however, the packets
would shrink or expand, in accordance with their (time-dependent)
semiclassical velocity.

Floquet engineering is a very effective method to stabilize wave
packets by means of a periodic drive, which does not need to be
strong. However, up to now it has only been applied to systems that
explicitly break spatial translation symmetry, such as electrons in
ionic potential, or atomic condensates in gravitational field. Here
the translation symmetry can be either continuous (for continuous
systems) or discrete (for lattice systems).  In this work, we
generalize the Floquet engineering approach to create nondispersive
wave packets in extended systems that do not break translation
symmetry in the bulk when undriven.  Examples include photons in
extended microwave resonators, or particles (atoms, electrons)
confined to a linear or circular resonators or chains of coupled
resonators \cite{schuster20,painter19,manucharyan19}. Surprisingly, we
discover that spatially localized driving is sufficient to generate
compact dispersionless wave packets that are Floquet eigenstates.  A
traveling wave packet on such a device could conceivably serve as a
``bus'', over which quantum information or a particle can be shuttled
across the entire chain.

\newcommand{\Trec}{T_{\text{rec}}}%
\newcommand{\Ttun}{T_{\text{tun}}}%

As a paradigmatic model we study a homogeneous tight binding chain.
Depending on the number of sites and energy, it can implement either
parabolic dispersion or approximate linear dispersion. The dispersion
slope determines the group velocity, $v_g$, which, in combination with
the system length, determines the wave packet's recurrence time
$\Trec$ (i.e., the time a wave packet takes to traverse one round trip
of the system). This recurrence time is in general a function of
energy.

A Floquet drive is defined by its period $T$, and its spatial and
temporal profiles.  %
We will see that $T$ singles out a series of spectral segments of the
undriven system that are most susceptible to the formation of wave
packets, as organized by their recurrence time $\Trec = \frac{s}{r}T$,
where $s$ and $r$ are co-prime integers.  The combination of the
drive's spatial and temporal profiles then imposes selection rules
that determine which wave packets actualize, as well as their
properties such as spatial compactness.  When $s>1$, the Floquet wave
packets manifestly break the discrete time-translation symmetry of the
drive; we will discuss the connection to time crystal physics
\cite{wilczek12,bruno13,watanabe15,sacha15,else16,khemani16,sacha17,zhang17}.
As long as these general rules are satisfied, the formation of wave
packets is robust with respect to details such as the overall drive
strength, the introduction of spatial or temporal randomness,
etc. This flexibility also opens up the ability to fine-tune drive
protocols for specific applications. As a proof of principle, we will
demonstrate how to design a drive that reproduces a particular target
Floquet micromotion.

\paragraph{Floquet wave packets at the primary resonance}
To build intuition, we first discuss the emergence of Floquet wave
packets at the primary resonance, that is those with a round trip time
equal to the drive period, $\Trec = T$. Consider a time-periodic
Hamiltonian $\op H(t+T) = \op H(t)$,
\begin{gather}
  \op H(t) = \op H_0 + \op V(t)\ , \ \op V(t) = \sum_a g_a \op V^{(a)} e^{ia\Omega t}\ ,
\end{gather}
where $\op H_0$ is the undriven tight-binding Hamiltonian, assumed to
be spatially homogeneous, $\op V^{(a)}$ encodes spatial dependence of
the drive at frequency $a\Omega$, with relative strength $g_a$, and
$\Omega = 2\pi/T$ is the fundamental frequency. %
After one drive period, a Floquet eigenstate $|\psi\rangle$ returns to
itself with an additional phase (\emph{quasienergy}),
$\op U_T|\psi\rangle = e^{-i\theta}|\psi\rangle$, where
$\op U_t = \mathcal{T}\exp \left[ -i\int_0^t dt' \op H(t') \right]$ is
the time evolution operator. This state can be lifted to a
time-periodic trajectory in Hilbert space, \emph{i.e.}, a
\emph{Floquet micromotion},
$|\psi(t)\rangle = |\psi(t+T)\rangle = e^{+i \frac{\theta}{T}t}\op
U_t|\psi\rangle$, which satisfies the Floquet-Schr\"odinger equation,
\begin{gather}
  \label{floq-sch}
  \left[ \op H(t) - i \partial_t  \right] |\psi(t)\rangle = \frac{\theta}{T}|\psi(t)\rangle\ .
\end{gather}
Note that shifting
$\theta \rightarrow \theta + 2\pi a \ (a \in \mathbb{Z})$ leads to
gauge equivalent micromotions
$|\psi(t)\rangle \rightarrow |\psi(t)\rangle e^{ia\Omega t}$ of the
same \emph{physical} time evolution. 

In the undriven limit, Floquet eigenstates are simply the energy
eigenstates $|\varepsilon_k\rangle$ of $\hat{H}_0$ with integer label
$k$.  At weak drive, thus, most Floquet eigenstates are close to an
undriven state and remain spatially extended. However, if the drive
frequency $\Omega = 2\pi / T$ is close to the level spacing $\Delta$
somewhere in the spectrum of $\op H_0$, then the drive can efficiently
couple several nearby unperturbed eigenstates. To describe this, one
can expand a generic dispersion relation around some $k_{*}$ (not
necessarily an integer), such that
\begin{gather}
  \label{ek-taylor}
  \varepsilon_k = \varepsilon_{{*}} + (k - k_{*})\Omega + \frac{u}{T}(k-k_{*})^2 + \cdots\ , \ k \in \mathbb{Z}\ ,
\end{gather}
and consider a micromotion \emph{ansatz}
\begin{gather}
  \label{psi-ansatz}
  |\psi(t)\rangle = \sum_k f_k|\varepsilon_k\rangle e^{-ik\Omega t}\ .
\end{gather}
Note that $|\varepsilon_k\rangle e^{-i k\Omega t}$ are Floquet
micromotions in the undriven limit, and the gauge ($a = k$) is chosen
so that near resonance, the corresponding quasienergies,
$\theta_k^{(0)} = \varepsilon_k T - 2\pi k$, are nearly
degenerate (in the scale of $\Omega T = 2\pi$), hence
Eq.~\ref{psi-ansatz} is akin to degenerate perturbation
solutions. For consistency, the range of the $k$ summation should be
constrained such that $\{\theta_k^{(0)}\}$ are roughly within a single
Floquet zone.

We assume positive $u$ and $\partial_k \varepsilon_k$ in
Eq.~\ref{ek-taylor}; the case with one or both of them negative can be
similarly handled. The coefficients $\{f_k\}$ are determined as the
``degenerate perturbation'' solutions of the Floquet-Schr\"odinger
operator in the Hilbert space of the ansatz (Eq.~\ref{psi-ansatz}),
\begin{gather}
  \label{eff1d}
  \sum_{k\pr} \left[ u (k-k_{*})^2 \delta_{k\,k\pr} + g_{k\pr - k} V_{kk\pr}\right] f_{k\pr} 
  = (\theta - \theta_{*}) f_k\ ,
\end{gather}
where
$V_{kk\pr} = \langle \varepsilon_k | \op V^{(k\pr - k)} |
\varepsilon_{k\pr}\rangle T$ and
$\theta_{*} = \varepsilon_{*} T - 2\pi k_{*}$. Eq.~\ref{eff1d} maps
our problem to an effective one-dimensional ``lattice'' with quadratic
``on-site potential'' $u(k-k_{*})^2$ and ``hopping''
$g_{k\pr - k}V_{kk\pr}$. One thus expects on general grounds that its
eigenstates will mix different $k$ ``sites.'' Translating back to the
original problem, the Floquet micromotion $|\psi(t)\rangle$ is thus a
linear superposition of momentum states $|\varepsilon_k\rangle$ with
time-independent weights $|f_k|^2$, and is therefore a wave packet in
coordinate space.

\begin{figure}
  \centering
  \includegraphics[width=.35\textwidth]{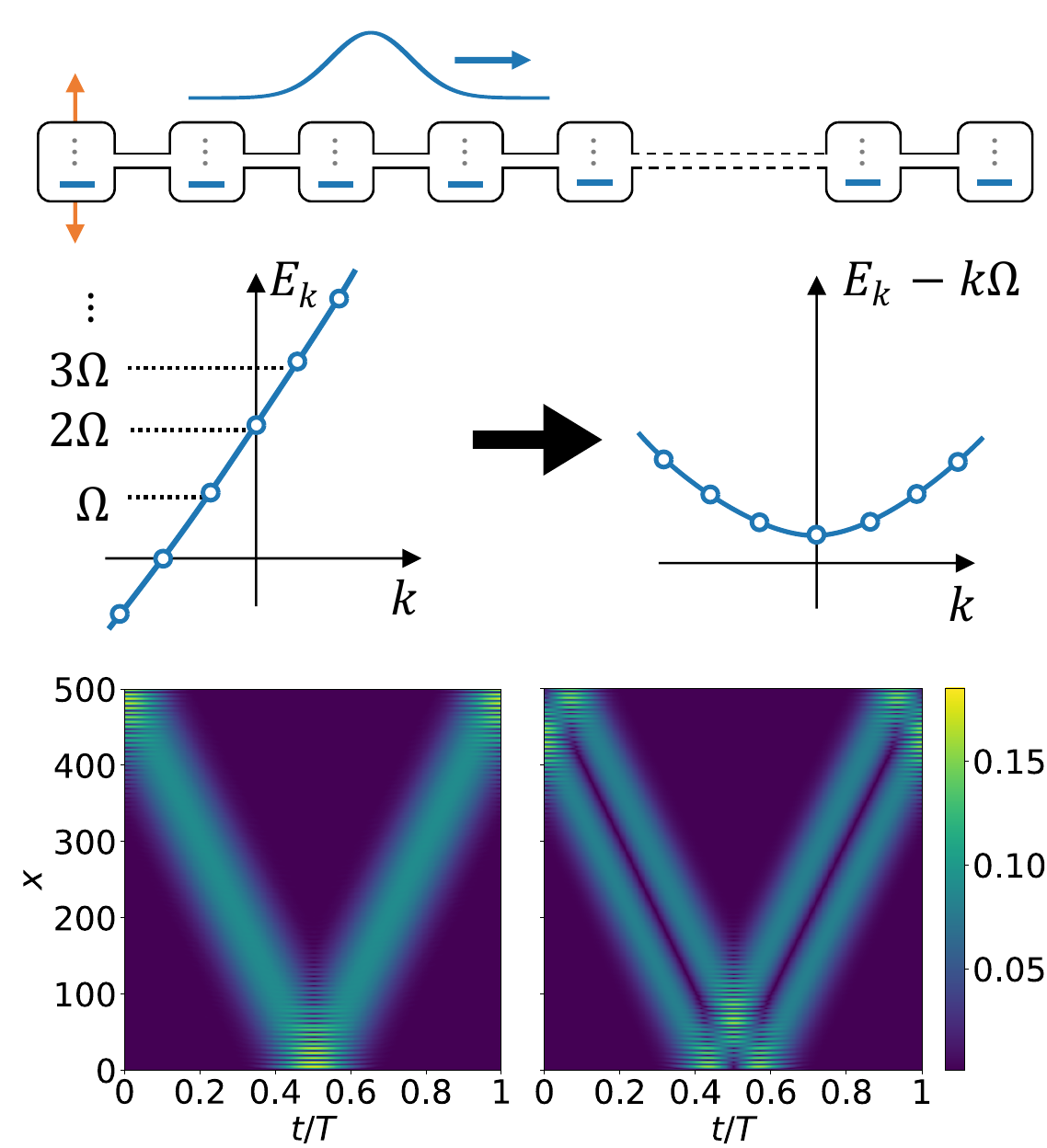}
  \caption{Top: schematic of coupled resonators.  Temporal modulation
    of the first site's onsite energy induces Floquet wave
    packets. Such a scenario is modeled by Eq.~\ref{hg}. Middle: At
    the primary resonance where the drive frequency matches the
    typical level spacing of the undriven problem (left), the
    effective model (Eq.~\ref{eff-sho}) is a lattice harmonic
    oscillator with quadratic on-site energy (right). Bottom: Two (of
    several) wave packet solutions of Eq.~\ref{hg} corresponding to
    the ground (left) and the first excited (right) states of the
    emergent lattice oscillator. The system size $L = 500$, drive
    strength $g = 1$, and drive period $T = 1005$. Note that the wave
    packets maintain their spatial compactness at all time.}
  \label{fig:sr1}
\end{figure}

As a concrete example, we consider an open boundary chain of length
$L$ driven on the first site
(Fig.~\ref{fig:sr1}),
\begin{gather}
  \label{hg}
  \op H(g) = \sum_{x=1}^{L-1}|x\rangle\langle x+1| + h.c. + 2 g \cos(\Omega t) |1\rangle\langle 1|\ ,
\end{gather}
where the only nonvanishing Fourier components of the drive are
$g_1 = g_{-1} = g$. This limits the effective hopping in
Eq.~\ref{eff1d} to nearest neighbor in $k$, and for simplicity, we
will approximate it as $k$-independent and evaluate it at $k_{*}$,
writing $\tau \equiv g V_{k_{*},k_{*}}$. Eq.~\ref{eff1d} then becomes
\begin{gather}
  \label{eff-sho}
  u(k-k_{*})^2 f_k + \tau (f_{k-1} + f_{k+1}) = (\theta - \theta_{*}) f_k\ ,
\end{gather}
and maps to a lattice version of harmonic oscillator, with stiffness
$u$ and hopping $\tau$. For sufficiently large $\tau/u$, a subset of
its eigenstates are thus Gaussian-like wave packets with
$1, 2, \cdots, D$ spatial peaks, where $D$ counts the number of
oscillator-like states. In Fig.~\ref{fig:sr1}, we plot two of the $D$
wave packet solutions corresponding to the ground and the first
excited states of Eq.~\ref{eff-sho} (and hence with one and two
spatial peaks, respectively). To form a wave packet, the ``hopping''
must be able to efficiently couple several $k$ states, hence $D$ can
be estimated as the number of ``sites'' that are energetically within
one hop's reach from the potential bottom,
$u \delta k^2 \le \tau\Rightarrow$ $|\delta k| \le \sqrt{\tau/u}$
$\Rightarrow D \simeq 2 \sqrt{\tau/u}$, where $\delta k = k -
k^{*}$. The crossover drive strength to induce any wave packet at all
is thus $\tau_c \simeq u/4$, although a substantially stronger drive
is needed to produce better spatial compactness (as $D$ also counts
the number of momentum constituents in a wave packet). The emergent
oscillator ``frequency,'' $\varpi = 2 \sqrt{\tau u}$, is approximately
the level spacing of the quasienergies $\{\theta\}$. Physically, thus,
if an initial state is a superposition of such wave-packet Floquet
eigenstates, it will (approximately) revive after $2\pi / \varpi$
drive periods. A locality-based measure, such as the participation
ratio $\sum_x|\psi(x,t)|^4$, will then exhibit beats at frequency
$\sim \varpi \Omega / (2\pi)$. %

\newcommand{\lp}{\mathcal{L}}%
To evaluate $u$ and $\tau$, we note that the undriven $\op H(0)$ has
eigenstates
$\langle x | \varepsilon_k\rangle = \sqrt{2/\lp}\sin(q_k x)$ with wave
vectors $q_k = \pi k/\lp$, and eigenvalues
$\varepsilon_k = 2\cos q_k$. Here $\lp \equiv L+1$ and
$x,k = 1, 2, \cdots, L$. From
$\partial_k \varepsilon_{k_{*}} = \Omega$ and
$\partial_k^2 \varepsilon_{k_{*}} = 2u/T$, we get
$u = \pi^2\sqrt{T^2 - \lp^2}/\lp^2$ and $\tau = 2g \lp/T$.
Parametrizing $\beta = \lp/T$ and $\gamma = 1/\sqrt{1-\beta^2}$, one
finds that the emergent ``frequency'' is
$\varpi = 2\pi \sqrt{2g /\gamma \lp}$, the number of Floquet wave
packets is $D \simeq \frac{2\beta}{\pi} \sqrt{2 g \gamma \lp}$, and
the crossover drive strength is $g_c = \pi^2 / 8\beta^2 \gamma \lp$.

A related semiclassical analysis was conducted in
Ref.~\cite{buchleitner02}. However, we point out that the results here
are more general and do not rely on the existence of a semiclassical
limit. An example of such (a ``cubic oscillator'') is provided in the
Supplemental Materials \cite{Note1}.

\paragraph{Floquet wave packets at rational resonances}
The same setup can more generally host many series of non-dispersing
wave packets that have recurrence times not just equal to, but
rationally commensurate with the drive period, $\Trec = \frac{s}{r}T$,
where $s,r$ are co-prime integers. The group velocity of an $(s,r)$
wave packet is $v_g = 2L / \Trec$ ($2L$ being the round trip length),
hence it consists mostly of states from the segment of the undriven
energy spectrum where the typical level spacing is
$\Delta = \frac{r}{s}\Omega$. We discuss here the more salient
features of such wave packet solutions, and leave mathematical details
to the SM \cite{Note1}.

\begin{figure}
  \centering
  \includegraphics[width=.35\textwidth]{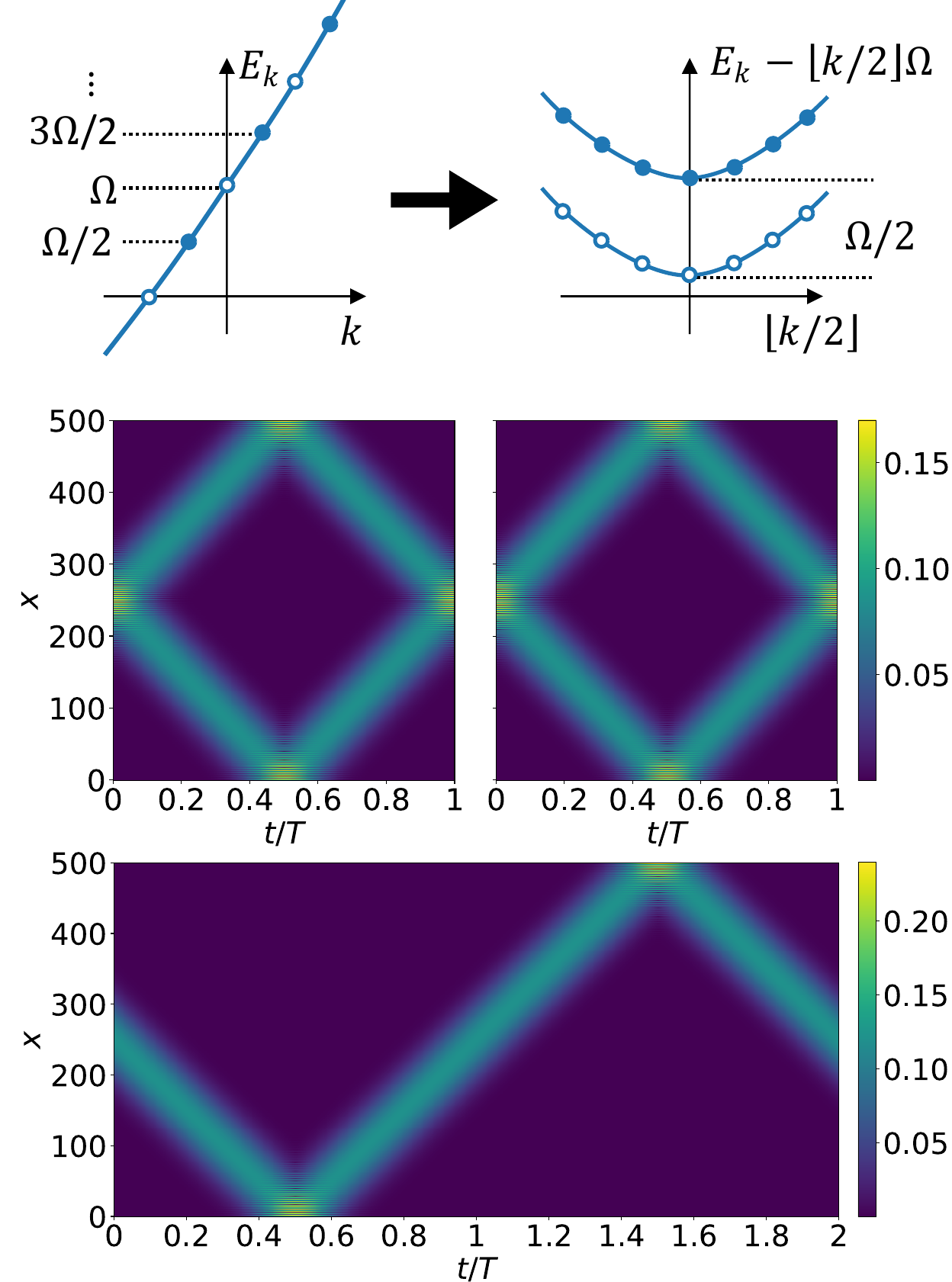}
  \caption{Top left: when the drive frequency matches twice the
    typical level spacing (i.e.~an $s=2,r=1$ resonance), the drive
    only couples $k$ of the same parity (solid or empty dots). The
    effective model becomes two independent chains with an overall
    $\Omega / 2$ gap between their ``onsite'' potentials (top
    right). Center: A twin pair of $s=2, r=1$ wave packet solutions of
    Eq.~\ref{hg} ($L=500, g = 1, T = 225$), corresponding to
    the respective ground states of the two effective chains. Their
    spatial-temporal patterns are almost indistinguishable. Their
    quasienergies are $\pi + \delta$ apart, where numerically
    $\delta \sim 2.6\times 10^{-6}\pi$. Bottom: dynamical evolution of
    a time-crystalline wave packet initialized as the sum of the two
    solutions, undergoing $r = 1$ round trip in $s = 2$ drive
    periods. After an very long tunneling time of
    $2\pi T / \delta \sim 7.6\times 10^5 T$, it would evolve into the
    wave packet configuration of the opposite linear combination
    (difference instead of sum).  }
  \label{fig:s2}
\end{figure}

For $s > 1, r = 1$, we consider $s = 2$ as a concrete example. To
leading order, the drive only resonantly couples within even
$k = 2\kappa$ and odd $k = 2\kappa + 1$, separately. One can thus use
an ansatz similar to Eq.~\ref{psi-ansatz} but restricted to a given
parity,
$|\psi^{(\sigma)}(t)\rangle = \sum_{\kappa}
f_{\kappa}^{\sigma}|\varepsilon_{2\kappa + \sigma}\rangle e^{-i\kappa
  \Omega t}$, where $\sigma = 0,1$ is the parity of $k$. Invoking
Eq.~\ref{floq-sch} then leads to two effective lattice models similar
to Eq.~\ref{eff1d}, one for each parity, see SM.  Similar to the
primary resonance case, one then concludes that wave packet solutions
generically exist above a crossover drive strength. Crucially, at
large $L$, the two effective chains are essentially identical except
for an overall $\frac{\Omega}{2}$ shift in the ``onsite'' energy
(Fig.~\ref{fig:s2} top). This translates to a $\pi$ gap between their
quasienergy spectra, and is the origin of time-crystalline nature of
individual wave packets, as we will see next.

In Fig.~\ref{fig:s2} (center), we plot two $(s,r) = (2,1)$ wave packet
solutions resulting from Eq.~\ref{hg}, which correspond to the
``ground state'' of the even- and odd-parity effective models,
respectively. As shown, both consist of two counter-propagating wave
packets that evolve into each other after one period. Thus, even
though the individual wave packet returns only after $2T$, the Floquet
eigenstate remains $T$-periodic. Individual wave packet can be
obtained by initializing into the sum (or difference) of the two
parity ground states. The evolution of one such combination is shown
in Fig.~\ref{fig:s2} (bottom). As mentioned before, the quasienergy
gap between the two parity-related states is $\pi + \delta$, where the
small deviation $\delta$ is due to higher order effects that mix the
two parity sectors. In the limit $\delta \rightarrow 0$, the
individual wave packets are perfectly stable, recurring after $2T$ --
a manifestation of time-translation symmetry breaking, analogous to
discrete time crystals.  A nonzero $\delta$ introduces a time scale
$2\pi T / \delta$, over which one time translation symmetry-broken
state tunnels into the other. We find numerically that this tunnelling
time can be indeed very long, reaching thousands of drive periods for
reasonable drive strengths and system sizes, and is easily
tunable. For Fig.~\ref{fig:s2}, the tunneling time is $\sim 10^5 T$.

\begin{figure}
  \centering
  \includegraphics[width=.38\textwidth]{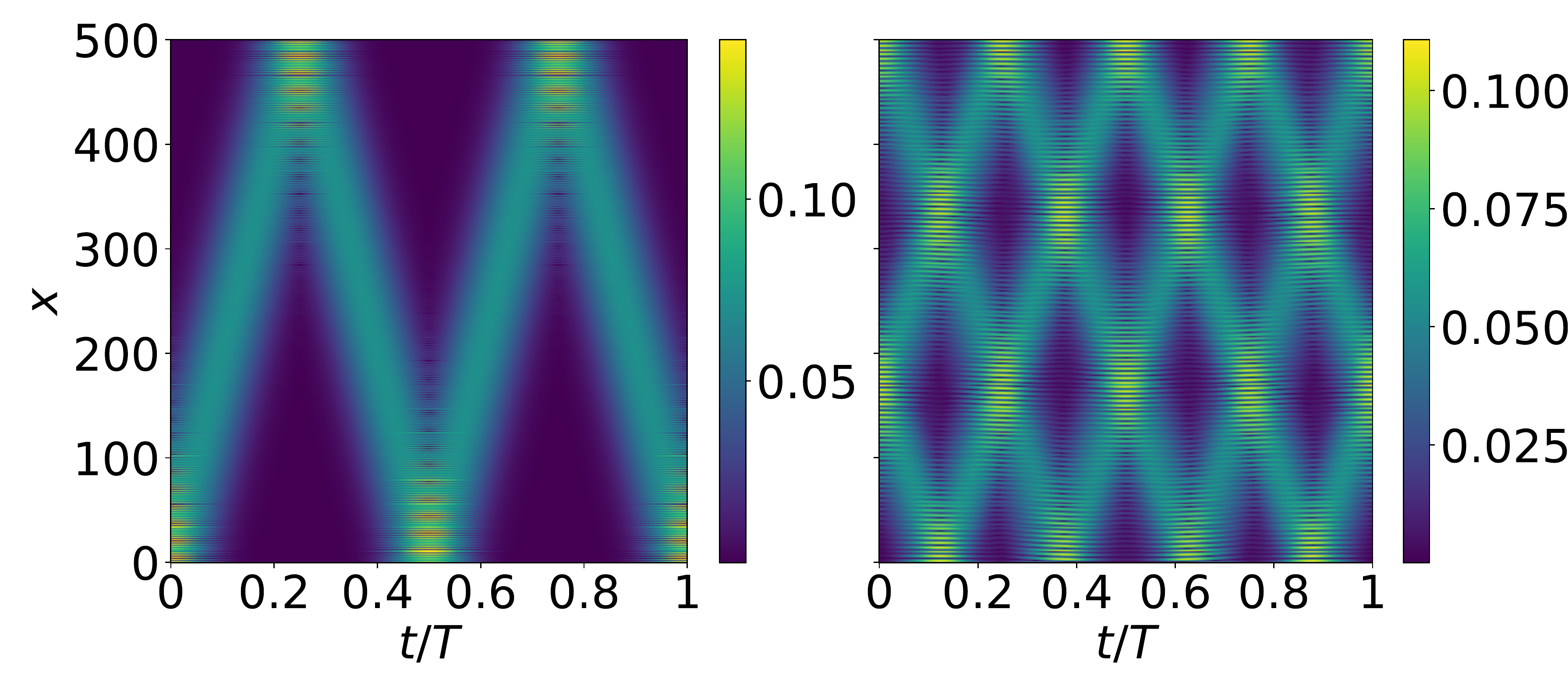}
  \caption{Floquet wave packets at generic $(s,r)$ resonances,
    \emph{coexisting} under the same drive Eq.~\ref{hg} with
    $L = 500, g = 1, T=1005$ (same as Fig.~\ref{fig:sr1}). Such
    Floquet states consist of $s$ traveling wave packets, each
    traversing $r$ round trips in $s$ drive periods. Left:
    $s = 1, r = 2$. Right: $s = 3, r = 4$.
  }
  \label{fig:r-mm}
\end{figure}

We note that time-crystallinity is typically discussed in the context
of interacting many body systems. The Floquet wave packets here are
single particle states, and their ``time crystallinity'' refers only
to their time-translation symmetry breaking behavior. Nevertheless, we
expect that such behavior would survive also in the many-body setting
in the presence of interaction; this will be a subject of future
work \cite{Note2}.

The analysis with $r > 1$ is technically more involved, and we leave
mathematical details to the SM, where we discuss a nontrivial
generalization of the ``degenerate perturbation'' ansatz
Eq.~\ref{psi-ansatz} and the resulting effective lattice model (a more
elaborate version of Eq.~\ref{eff-sho}). We find that an $(s,r)$
Floquet eigenstate consists of $s$ wave packets, each completing a
fraction $\frac{r}{s}$ of round trip in one drive period, see
Fig.~\ref{fig:r-mm}. Like the $s=2$ case discussed before, a given
$(s,r)$ solution is one of $s$ partners with almost identical
spatial-temporal patterns, and their quasienergies are equally spaced
by $\Delta\theta = 2\pi/s$ to leading order. The individual wave
packets can be resolved by linear combinations of the $s$ partners,
hence their true recurrence time is $2\pi / \Delta \theta =
sT$. However, since they completed $r$ round trips in $sT$, their
\emph{apparent} recurrence time is $\Trec = sT/r$. The rational ratio
of $\Trec/T$ is suggestive of a \emph{fractional time crystalline
  order}, a notion put forward very recently \cite{matus19, pizzi19}.

\newcommand{\psitarg}{\tilde\psi}
\newcommand{\ftarg}{\tilde f}
\paragraph{Floquet drive engineering}
Finally, we discuss how to realize a desired target micromotion
through drive engineering.  Assume the time-dependent Hamiltonian has
a form $\op H(t) = \sum_n w_n Q_n(t) + h.c.$ where
$Q_n = \op h_n e^{i a_n \Omega t}$ represent experimentally available
Hamiltonian controls $\op h_n$ at integer harmonics $a_n$, and $w_n$
are (generally) complex-valued coefficients. Given a target
micromotion $|\psitarg(t)\rangle$ and a prescribed set of $\{Q_n\}$,
one can ask what is the best choice of $\{w_n\}$ to produce a
micromotion as close to the target as possible. Writing the
Floquet-Schr\"odinger operator as $K = \op H(t) - Q_0$ where
$Q_0 = i\partial_t$, the optimal \newcommand{\ee}{\textsf{E}}
coefficients are those that minimize the variance
$\Delta = \langle K^2\rangle_c$, where
$\langle \mathcal{O} \mathcal{O}\pr \rangle_c = \langle \mathcal{O}
\mathcal{O}\pr \rangle - \langle \mathcal{O}\rangle\langle
\mathcal{O}\pr\rangle$ and
$\langle \mathcal{O}\rangle = \int_0^T dt \langle \psitarg |
\mathcal{O} | \psitarg\rangle$.  By construction, $\Delta \ge 0$ and
vanishes only if $|\psitarg\rangle$ is an exact eigenstate of $K$
\cite{qi17,chertkov18,greiter18}. Demanding
$0 = \frac{\partial \Delta}{\partial w_n} = \frac{\partial \Delta}{
  \partial w_n^{*}}$ then yields the solution
$\smat{ \vec w \\ \vec w^{*}} = \smat{G & F \\ F^{*} & G^{*}}^{-1}
\smat{\vec J \\ \vec J^{*}}$, where
$G_{mn} = \langle Q_m Q_n + Q_n Q_m\rangle_c$,
$F_{mn} = \langle Q_m Q_n\dg + Q_n\dg Q_m + h.c.\rangle_c$,
$J_n = \langle Q_0 Q_n + Q_n Q_0\rangle_c$, and $(\cdot)^{-1}$ is the
Moore-Penrose pseudo-inverse. As a proof of principle, we target a
non-dispersing Gaussian wave packet,
$|\psitarg(t)\rangle = \sum_k \ftarg_k |k\rangle e^{-i k \Omega t}$
where $\ftarg_k \propto e^{-(k-k_0)/4\sigma^2}$. On a chain of length
$L = 100$, for example, we can realize this wave packet as a Floquet
eigenstate (to a high fidelity of $ > 0.99$ at all time) using only
on-site drives and only two frequencies (i.e.~the first and second
harmonic); see Fig.~\ref{fig:parent}. In contrast, the static
Hamiltonian necessary to sustain such a dynamically non-dispersing
wave packet, $\sum_k |k\rangle k\Omega \langle k|$, is spatially
highly nonlocal. Note that targeting a different micromotion (e.g.,
one with different $k_0$ and $\sigma$) generally results in a
different optimal drive.  Fidelity with the target state can be
further enhanced with more drive terms such as local hops or higher
harmonic modulations. Additional requirements such as spatial
smoothness of the drive can be implemented by including corresponding
penalty terms in $\Delta$.

\begin{figure}
  \centering
  \includegraphics[width=.5\textwidth]{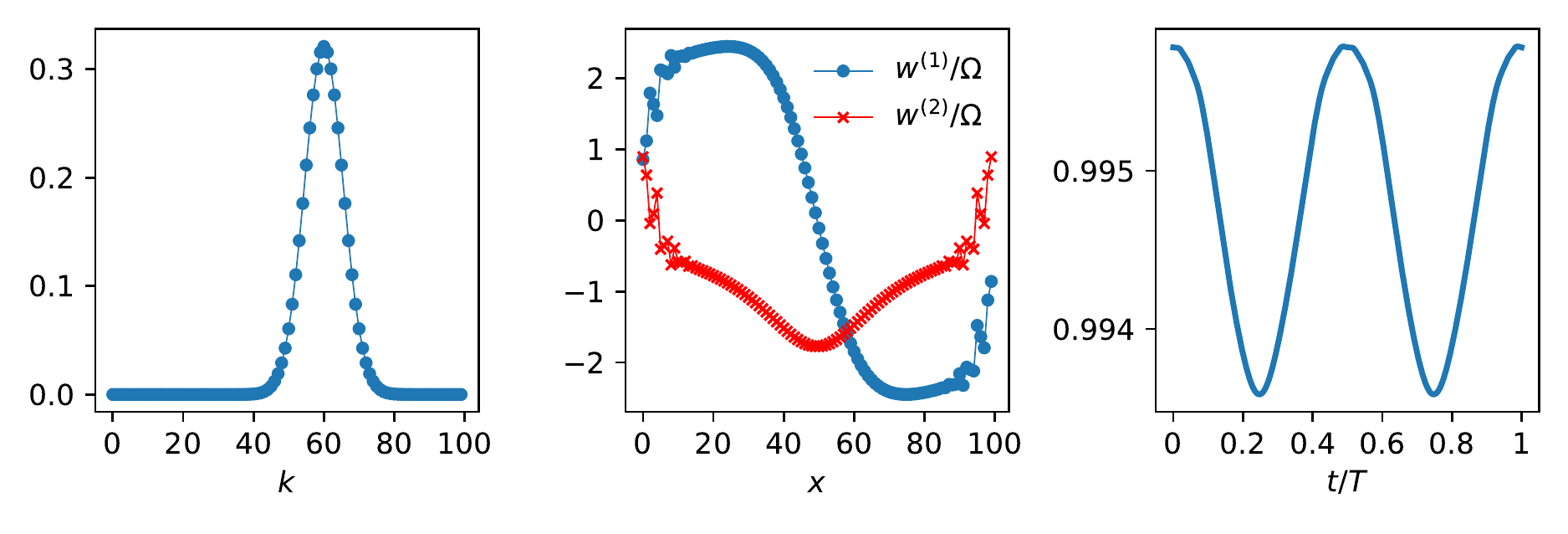}  
  \caption{Optimal drive to reproduce a target micromotion
    $|\psitarg(t)\rangle = \sum_k e^{-(k-k_0)/4\sigma^2 -ik\Omega
      t}|\varepsilon_k\rangle$ on a chain of length $L=100$, with
    $k_0 = 60$ and $\sigma^2 = 15$. We use static,
    translation-invariant nearest neighbor hopping, and onsite
    modulation up to the second harmonic,
    $\op H(t) = \sum_x w^{(0)} |x\rangle\langle x + 1| + [ w_x^{(1)}
    e^{i\Omega t} + w_x^{(2)} e^{2i\Omega t} ]|x\rangle\langle x| +
    h.c$. Left: $k$-space Gaussian profile of the target
    state. Center: optimal drive strengths; the optimal static hopping
    is $w^{(0)} / \Omega = 17.09$. Note that all $w$ numerically turn
    out to be real-valued even though they are allowed to be
    complex. Right: Fidelity $|\langle \psitarg | \psi\rangle|$
    between reproduced and target micromotions over one period. }
  \label{fig:parent}
\end{figure}

\paragraph{Discussion}
We note in closing that effective models like Eq.~\ref{eff1d} allow us
to efficiently reason about more complicated drives.  For example, in
Eq.~\ref{hg}, while keeping the temporal profile as $\cos(\Omega t)$,
one could replace the single site modulation with
$\sum_x v(x)|x\rangle\langle x|$ of an arbitrary---potentially fully
random---spatial profile $v(x)$, yet the product form
$g_{k\pr - k}V_{kk\pr}$ in Eq.~\ref{eff1d} automatically filters out
all but one Fourier component in $v(x)$, hence its only effect is to
renormalize the drive strength. This implies, among other things, that
the resulting wave packets do not perceive any spatial randomness in
the drive.  On the other hand, if one fine-tunes $v(x)$ such that a
particular spatial Fourier component vanishes exactly, then the
corresponding resonance will be fully suppressed.

The compact wave packets reported here are closely related to
spatio-temporal focusing that leads to pulse formation in long
parametrically modulated resonators, such as mode-locked lasers. In
the case of perfectly linear dispersion (as for electromagnetic waves
in vacuum), parametric pumping leads to exponential amplification and
spatial compression of any initial field configuration
\cite{martin2019floquet}; in the tight binding model discussed here,
the exponential explosion is truncated due to the energy dependence of
group velocity, leading instead to formation of compact travelling
Floquet eigenstates.

\paragraph{Acknowledgment}
We are grateful to Ian Mondragon-Shem, D. Schuster, V. Manucharyan,
A. V. Balatsky, A. Saxena, and D. P. Arovas for useful discussions and
feedbacks. This material is based upon work supported by Laboratory
Directed Research and Development (LDRD) funding from Argonne National
Laboratory, provided by the Director, Office of Science, of the
U.S. Department of Energy under Contract No. DE-AC02-06CH11357

\bibliography{fwp}

\clearpage
\onecolumngrid
\appendix
\section{Supplemental Materials}
In this note, we provide details on the analysis of $r \neq s \neq 1$
Floquet wave packets. We first discuss analytically tractable cases
where one of $r$ and $s$ is $1$. When neither of them is $1$, an
effective lattice model can still be derived, although it does not
yield to analytical solution, and we discuss its qualitative
features. We then briefly discuss the special case where the Floquet
drive resonates with the undriven energy spectrum close to its
inflection point, which leads to an effective lattice model with a
cubic ``potential''. Finally we estimate the change in the width of
the wave packets during one period.

\section{Effective model for $s > 1, r=1$}
In this section, we discuss the effective model for the $s > 1, r=1$
resonance, where the drive frequency matches $s$ times the typical
level spacing, $\Omega \simeq s \Delta$. A generic undriven energy
spectrum can be expanded as ($k_{*}$ not integer in general)
\begin{gather}
  \varepsilon_k = \varepsilon_{k_{*}} + (k-k_{*}) \frac{\Omega}{s} + \frac{u}{T}(k-k_{*})^2 + \cdots\ .
\end{gather}
To leading order, the drive only resonantly couples level $k$ to
$k \pm s$. This effectively separates the undriven energy eigenstates
into $s$ subspaces according to $\sigma = k \mod s$. For example, when
$s = 2$, $\sigma = 0,1$ is the parity of $k$, and to leading order,
the drive does not mix states of different parity. Writing
\begin{gather}
  k = \kappa s + \sigma\ ,
\end{gather}
then within each subspace $\sigma$, the integer $\kappa$ plays the
role of $k$ in primary resonance, hence we can use an ansatz
\begin{gather}
  \label{ansatz-s}
  |\psi^{\sigma}(t)\rangle = \sum_{\kappa} f_{\kappa}^{\sigma}|\varepsilon_{\kappa,\sigma}\rangle e^{-i\kappa \Omega t}\ .
\end{gather}
Note that when $s = 1$, $\sigma$ can only be $0$, and the ansatz above
reduces to that of the primary resonance discussed in the text. Recall
that the Floquet-Schrodinger equation is
\begin{gather}
  \label{fs}
  \left[ \op H(t) - i\partial_t \right]|\psi(t)\rangle =
  \frac{\theta}{T}|\psi(t)\rangle\ ,
\end{gather}
where the time-dependent Hamiltonian is
\begin{gather}
  \op H(t) = \op H_0 + \op V(t)\quad , \quad \op V(t) = \sum_a g_a \op V^{(a)}e^{ia\Omega t}\ .
\end{gather}
Solving Eq.~\ref{fs} with \ref{ansatz-s} then yields an eigenvalue equation
\begin{gather}
  \label{eff-s}
  \sum_{\kappa\pr} \left[ s^2 u(\kappa - \kappa_{*}^{\sigma})^2 \delta_{\kappa \kappa\pr} + g_{\kappa\pr - \kappa} V^{\sigma}_{\kappa\kappa\pr}\right] f_{\kappa\pr}^{\sigma} = (\theta^{\sigma} - \theta^{\sigma}_{*}) f_{\kappa}^{\sigma}\ ,
\end{gather}
where
\begin{gather}
  V_{\kappa\kappa\pr}^{\sigma} = \langle \varepsilon_{\kappa,\sigma} |\op V^{(\kappa\pr - \kappa)}|\varepsilon_{\kappa\pr,\sigma}\rangle T\quad , \quad \kappa_{*}^{\sigma} = \frac{k_{*} - \sigma}{s}\quad , \quad \theta^{\sigma}_{*} = \varepsilon_{k_{*}} T - 2\pi \kappa_{*}^{\sigma}\ .
\end{gather}
The effective model, Eq.~\ref{eff-s}, thus consists of $s$ independent
``chains'', where $\sigma$ labels the chains, and $\kappa$ labels
``sites'' within each chain. The ``onsite potential'' is quadratic,
$u(\kappa - \kappa_{*}^{\sigma})^2$, and each chain has its own
quasienergy shift (i.e., a chain-dependent ``chemical potential'')
$\theta_{*}^{\sigma}$.

In a large system with $L$ physical sites,
$V^{\sigma}_{\kappa \kappa\pr}$ becomes $\sigma$-independent (The
leading order correction due to finite $L$ is $\sim L^{-1}$. E.g., in
an open boundary chain, it comes from
$ \delta q_k \frac{\partial}{\partial q_k} |\varepsilon_k\rangle$
where $q_k = \frac{k\pi}{L+1}$ is the wave vector in an open boundary
chain, and $\delta q_k = q_{k+1} - q_k \propto L^{-1}$). Hence
Eq.~\ref{eff-s} for different $\sigma$ have the same set of
eigenvalues $\{\theta^{\sigma} - \theta^{\sigma}_{*}\}$. The
quasienergies $\{\theta^{\sigma}\}$ from different chains $\sigma$ are
thus ``gapped'' from each other by
$\theta^{\sigma+1}_{*} - \theta^{\sigma}_{*} = \frac{2\pi}{s}$, but
otherwise identical. In other words, the $i^{th}$ quasienergy on
``chain'' $\sigma$ is
$\theta_i^{\sigma} = \theta_i^0 + \frac{2\pi \sigma}{s}$, where
$\theta_i^0$ is the $i^{th}$ quasienergy on ``chain'' $\sigma =
0$. Thus with the same index $i$, there are $s$ Floquet eigenstates
with different $\sigma$ labels whose quasienergies are equally spaced
by $\Delta \theta = 2\pi / s$. The time evolution of an arbitrary
linear superposition of these Floquet eigenstates thus have a
recurrence time of $2\pi T/\Delta \theta = s T$. Such recombined
states manifestly break the time translation symmetry of the driving
Hamiltonian, which is periodic in $T$, and are thus single particle
analogues of discrete time crystals.

Let us now discuss the spatial feature of these Floquet eigenstates
and their time-crystalline linear recombinations, assuming the
undriven states are momentum eigenstates of an open boundary chain,
$\langle x | \varepsilon_k\rangle = \sqrt{\frac{2}{L+1}}\sin(q_k x)$
where the wave vectors are $q_k = k \frac{\pi}{L+1}$.  A Floquet
eigenstate $|\psi^{\sigma}\rangle$ of a given $\sigma$
(Eq.~\ref{ansatz-s}) is a linear combination of momentum states with
the same $\sigma$, and are therefore invariant under spatial
translation by $2L/s$ (with phase shift $2\pi \sigma / s$), where the
system size $L$ is half the round trip length. To conform with this
translation symmetry, $|\psi^{\sigma}\rangle$ for any $\sigma$ must
consist of $s$ spatial packets equally spaced along the round trip. To
resolve these spatial packets, we Fourier transform the set of
$\{|\psi^{\sigma}\rangle\}$ states at $t=0$,
\begin{gather}
  \label{phi-resolv}
  |\phi^{\lambda}\rangle = \sum_{\sigma = 0}^{s-1} e^{-i2\pi \lambda \sigma/s} |\psi^{\sigma}(0)\rangle\quad , \quad \lambda = 0, 1, \cdots, s-1\ .
\end{gather}
As discussed before, in the limit where the level spacing of the
quasienergies $\Delta \theta = 2\pi / s$ is exact (i.e., when (1) we
ignore higher order effect of the drive that mixes different $\sigma$
sectors, and (2) $V^{\sigma}_{\kappa\kappa\pr}$ becomes
$\sigma$-independent at large $L$), any linear combination of
$\{|\psi^{\sigma}\rangle\}$ breaks the discrete time translation
symmetry of the driving Hamiltonian. For $|\phi^{\lambda}\rangle$, we
have
\begin{gather}
  \op U_{T}|\phi^{\lambda}\rangle =
  e^{-i\theta^0}|\phi^{\lambda+1}\rangle\implies
  \op U_{sT}|\phi^{\lambda}\rangle =
  e^{-is\theta^0}|\phi^{\lambda}\rangle\ ,
\end{gather}
where $\op U_{sT}$ is dynamical time evolution over $s$ drive
periods. In other words, the $|\phi^{\lambda}\rangle$ states evolve
into each other after one $T$, and recur after $sT$. Physically, each
$|\phi^{\lambda}\rangle$ corresponds to a \emph{single} spatial packet
that propagates by $2L/s$ after $T$, and completes a round trip of
length $2L$ after $sT$.

Numerically, the quasienergy spacing among the $s$ partners
$\{|\psi^{\sigma}\rangle\}$ is $\Delta \theta = 2\pi / s + \delta$,
where a small $\delta$ originates from higher order effect of the
drive that mixes different $\sigma$ sectors, as well as the $\sigma$
dependence in $V^{\sigma}_{\kappa\kappa\pr}$. Consequently, the
$|\phi^{\lambda}\rangle$ states will ``tunnel'' among the $s$ wave
packet configurations over a time scale of $2\pi
T/\delta$. Numerically, the tunneling time is typically of the order
of thousands of drive periods, and may be extended further via
parameter fine tuning.

When $\op V(t) = 2g \cos \Omega t | 1\rangle\langle 1|$, i.e.,
a modulation on the first site at the fundamental frequency, the
effective model Eq.~\ref{eff-s} of a given $\sigma$ becomes a lattice version of harmonic oscillator,
\begin{gather}
  s^2u(\kappa - \kappa_{*}^{\sigma})^2 f_{\kappa}^{\sigma} + \tau (f_{\kappa - 1}^{\sigma} + f_{\kappa + 1}^{\sigma}) = (\theta^{\sigma} - \theta^{\sigma}_{*}) f_{\kappa}^{\sigma}\ ,
\end{gather}
where the parameters $u$ and $\tau$ can be estimated using $\partial_k \varepsilon_{k_{*}} = \Omega/s$ and $\partial^2_k \varepsilon_{k_*} = 2s^2 u/T$,
\begin{gather}
  u = \frac{\pi^2}{\lp^2} \sqrt{T^2 - \lp^2/s^2}\quad , \quad \tau = \frac{2g}{s^2} \frac{\lp}{T} \ .
\end{gather}
Here $\lp = L+1$. Note that the effective stiffness is now $s^2
u$. Parametrizing
\begin{gather}
  \beta_s = \frac{\lp}{sT}\quad , \quad \gamma_s = \frac{1}{\sqrt{1-\beta_s^2}}\ ,
\end{gather}
then similar to the primary resonance case, one can estimate the
emergent oscillator ``frequency'' $\varpi_s$ and the number of wave
packet solutions (per $\sigma$) $D_s$ as
\begin{gather}
  \label{efreq-d-s}
  \varpi_s = 2 \sqrt{s^2 u\tau} = 2\pi \sqrt{\frac{2g}{\gamma_s \lp}}\quad , \quad D_s \simeq 2 \sqrt{\frac{\tau}{s^2 u}} = \frac{2\beta_s}{s\pi} \sqrt{2g \gamma_s \lp}\ .
\end{gather}
These reduce to the primary resonance results of the main text when
$s = 1$. The crossover drive strength $g_c^{(s)}$ to induce any
$s > 1, r=1$ wave packet solution at all is
\begin{gather}
  D_s(g_c^{(s)}) = 1 \implies g_c^{(s)} = \frac{s^2\pi^2}{8\beta_s^2
    \gamma_s \lp}\ .
\end{gather}
Thus one generally needs a stronger drive to induce wave packets of
larger $s$.

\section{Effective model for $r > 1, s = 1$}
The situation with $r > 1$ is more involved. Consider first $s = 1$,
then it takes $r$ drive quanta at frequency $\Omega$ to resonantly
connect two adjacent energy levels, as they have a spacing
$\sim r\Omega$. As a result, a ``degenerate perturbation'' ansatz
similar to Eq.~4 in the main text would not work: the
Floquet-Schr\"odinger operator simply does not have matrix element
between $|\varepsilon_k\rangle e^{-irk\Omega t}$ and
$|\varepsilon_{k+1}\rangle e^{-ir(k+1)\Omega t}$. In principle, one
could attempt to derive an effective coupling between these levels via
an $r^{th}$ order perturbation theory; this is however technically
unwieldy.

We instead take an alternate route.  We are interested in wave packets
which traverse $r$ round trips of an $L$-site system in a single drive
period. Heuristically, this can be ``unfolded'' into one round trip in
a system of length $rL$---much like how the trajectory of a billiard
ball bouncing off the pool table can be ``unfolded'' into a straight
line across a repetitive tile of tables. This suggests that a proper
ansatz should additionally include eigenstates of the unfolded system,
\emph{truncated} to a segment of length $L$. These correspond to
fractional momentum states $|k + \frac{\rho}{r}\rangle$ in the
original system ($\rho = 0, 1, \cdots, r-1$); for the open chain
considered before,
$\langle x | k+ \frac{\rho}{r}\rangle \propto
\sin(q_{k+\frac{\rho}{r}}x + \varphi)$, where $\varphi$ is a phase
shift depending on how the shorter system is embedded into the longer
one.  For drives localized on $x=1$, as we will show, $\varphi$ is
such that $\langle L+1|k+\frac{\rho}{r}\rangle = 0$.

In the remainder of this section, we first justify the use of
fractional momentum states from perturbation theory, and then analyze
a generalized ansatz that additionally includes these states.

\subsection{The origin of fractional momentum states}
We argued that when the drive frequency matches $1/r$ of typical level
spacing of a tight binding chain of length $L$, the ansatz for Floquet
eigenstates should additionally include fractional momentum states,
which are energy eigenstates \emph{not} of a system of length $L$, but
rather of length $rL$. We now show that such fractional momentum
states do emerge as the leading order correction to undriven Floquet
eigenstates (the integer momentum states) when the Floquet drive is
treated as a perturbation. From the perspective of variational
solutions, thus, the purpose of including fractional momentum states
in the generalized ansatz is so that the variational subspace remains
invariant (to leading order) upon the action of the drive.

We first note that the Floquet-Schrodinger operator,
\begin{gather}
  K = K_0 + \op V(t)\quad , \quad K_0 = \op H_0 - i\partial_t\ ,
\end{gather}
acts on the tensor product space of the physical Hilbert space and the
space of periodic functions. The eigenvectors of $K_0$ (which are
space-time modes) form a complete basis in this space,
\begin{gather}
  |k,a\rangle \equiv |\varepsilon_k\rangle e^{-i a \Omega t}\quad , \quad   K_0 |k,a\rangle = \varepsilon_k - a\Omega\ ,
\end{gather}
where $|\varepsilon_k\rangle$ are eigenstates of the undriven
Hamiltonian, $\op H_0 = \sum_{x=1}^{L-1}|x\rangle\langle x+1| + h.c.$,
\begin{gather}
  \op H_0 |\varepsilon_k \rangle = \varepsilon_k |\varepsilon_k\rangle\quad , \quad   \varepsilon_k = 2 \cos q_k\ , \\
  \label{ek}
  \langle x|\varepsilon_k\rangle = \sqrt{\frac{2}{\mathcal{L}}} \sin(q_k x)\quad , \quad q_k = k\frac{\pi}{\mathcal{L}}\quad , \quad
  k,x = 1, 2, \cdots, L\quad , \quad \mathcal{L} = L+1\ .
\end{gather}
Note that the frequency index $a$ in $|k,a\rangle$ is independent of
the momentum index $k$. This is unlike the ansatz we used in the main
text, which associates to each momentum index a specific frequency
index (e.g., $a = k$ for primary resonance) --- that is, the ansatz
amounts to a variational solution in a subspace (of the full tensor
product space) in which frequency and momentum \emph{are} correlated.

Given an operator $Q = Q_0 + Q_1$, and unperturbed basis $|n\rangle$
with $Q_0 |n\rangle = \lambda_n |n\rangle$, the first order correction
to the eigenvectors $|n\rangle$ are
\begin{gather}
  |n^{(1)}\rangle = \sum_{m \neq n} \frac{\langle m | Q_1 | n\rangle}{\lambda_n - \lambda_m}|m\rangle\ .
\end{gather}
Now consider a Floquet drive
\begin{gather}
  \op V(t) = 2g \cos(\Omega t) \op v\ ,
\end{gather}
Treating $\op V(t)$ as a perturbation to $K_0$, we obtain the first
order correction to the undriven modes as
\begin{gather}
  \notag
  |k,a^{(1)}\rangle = g \sum_{\substack{(k\pr,a\pr) \\ \neq (k,a)}} \frac{\langle \varepsilon_{k\pr} | \op v | \varepsilon_k\rangle \langle a\pr | 2\cos(\Omega t) | a\rangle}{\varepsilon_k - \varepsilon_{k\pr} + (a-a\pr)\Omega} |k\pr,a\pr\rangle
                    =\check  g \sum_{\rho = \pm 1} |\chi_{k,\rho}\rangle e^{-i(a+\rho) \Omega t}\ , \\
|\chi_{k,\rho}\rangle = \sum_{k\pr}\frac{\langle \varepsilon_{k\pr} | \op v | \varepsilon_k\rangle}{(\varepsilon_k - \rho \Omega) - \varepsilon_{k\pr}}|k\pr\rangle\ ,
\end{gather}
where
$\langle a\pr | f(t) | a\rangle = \int_0^T \frac{dt}{T}\, e^{i(a\pr -
  a) \Omega t}f(t)$. Since we are considering a near resonance where
the drive frequency $\Omega$ matches $1/r$ of typical level spacing,
$\varepsilon_k - \rho \Omega$ is roughly the interpolation of the
dispersion relation $\varepsilon_k = 2\cos q_k$ at a fractional momentum
$\kappa_{k,\rho}$,
\begin{gather}
  \varepsilon_k - \rho \Omega \simeq \varepsilon_{\kappa_{k,\rho}}\quad , \quad \kappa_{k,\rho} = k + \frac{\rho}{r}
\end{gather}

Let us specialize to $\op v = |L\rangle\langle L|$, i.e., a drive on
the \emph{last} site on the chain, instead of the first site (as used
in the main text). This choice is for notational convenience only, and
we will comment on what changes if the drive is placed on the first
site later. Using
$\langle \varepsilon_{k\pr} | \op v | \varepsilon_k\rangle =
\frac{2}{\mathcal{L}}(-1)^{k+k\pr}\sin q_k \sin q_{k\pr}$, we have
\begin{gather}
  \label{chi}
  |\chi_{k,\rho}\rangle = (-1)^{k+1} \sin q_k \left[
    \frac{2}{\mathcal{L}}(-1)^{k\pr} \sum_{k\pr} \frac{\sin
      q_{k\pr}}{\varepsilon_{k\pr} - \varepsilon_{\kappa_{\rho}}} |k\pr\rangle\right]\ .
\end{gather}

We now show that $|\chi_{k,\rho}\rangle$ are indeed proportional to
fractional momentum states $|\kappa\rangle$, which are defined as the
interpolation of the integer momentum states (Eq.~\ref{ek}) to
non-integer momentum ``index'' $\kappa$,
\begin{gather}
\label{frac-k}
  \langle x | \kappa\rangle = \sqrt{\frac{2}{\mathcal{L}}}
  \sin(q_{\kappa} x)\quad , \quad q_{\kappa} = \kappa
  \frac{\pi}{\mathcal{L}}\  \forall \kappa\ .
\end{gather}
The overlap of two such states is
\begin{gather}
  \langle \kappa\pr | \kappa\rangle = I(\kappa - \kappa\pr) - I(\kappa + \kappa\pr)\ , \\
  I(\eta) \equiv \frac{1}{\mathcal{L}}\sum_{x=1}^L \cos \frac{\eta \pi x}{\mathcal{L}} = \frac{1}{2 \mathcal{L}} \left[ \sin(\eta \pi)\cot \frac{\eta\pi}{2 \mathcal{L}} - \cos(\eta \pi) - 1 \right]\ .
\end{gather}
Setting $\kappa\pr$ to integer  yields the expansion of $|\kappa\rangle$ in the integer momentum basis,
\begin{gather}
  \langle \varepsilon_k | \kappa\rangle = \frac{(-1)^k}{\mathcal{L}} \frac{\sin(\kappa \pi)\sin q_k}{\cos q_k - \cos q_{\kappa}}\ .
\end{gather}
Comparing with Eq.~\ref{chi} and noting that
$\cos q_{\kappa} = \frac{1}{2}\varepsilon_{\kappa}$, we find that
indeed $|\chi_{\kappa,\rho}\rangle$ are fractional momentum states,
\begin{gather}
  |\chi_{k,\rho}\rangle = (-1)^{k+1} \frac{\sin q_k}{\sin(\kappa_{k,\rho}\pi)}|\kappa_{k,\rho}\rangle\ .
\end{gather}
The effect of the Floquet drive on the undriven modes
$|k,a\rangle = |\varepsilon_k\rangle e^{-ia \Omega t}$ is thus to
bring an integer momentum state $|k\rangle$ at frequency $a$ to
fractional momenta $|k \pm \frac{1}{r}\rangle$ at neighboring
frequencies $a \pm 1$.

Note that in the case of primary resonance, viz.,
  $r = \rho = 1$, the drive brings $|k,a\rangle$ to
  $|k \pm 1, a \pm 1\rangle$. In other words, the $k=a$ subspace
  remains invariant under the drive, which justifies the ansatz used
  in the text,
  $|\psi(t)\rangle = \sum_k f_k |\varepsilon_k\rangle e^{-ik\Omega t}$.

What if we place the drive on the first site instead of the last one?
This is equivalent to relabeling site $x$ to $L+1-x$, hence the
appropriate fractional momentum states $|\wt \kappa\rangle$ are
related to $|\kappa\rangle$ (the ones arising from a last site drive)
by $\langle x | \wt \kappa\rangle = \langle L+1-x |
\kappa\rangle$. This effectively shifts $|\wt \kappa\rangle$ to a
different boundary condition,
$\langle x | \wt \kappa\rangle\propto \sin(q_{\kappa} x + \varphi)$
where $\varphi$ is such that $\langle L+1 | \wt \kappa\rangle = 0$. 

\subsection{Generalized ansatz and effective model}
We now discuss the effective model for the $r>1, s=1$ resonance, where
the drive frequency matches a fraction of the typical level spacing,
$\Omega \simeq \frac{\Delta}{r}$. From a group velocity consideration,
in one drive period, a wave packet consisting of states from this part
of the undriven spectrum (assuming it can be stabilized) will undergo
$r$ round trips (i.e., $2rL$ for an open chain of length $L$). Earlier
in this section, we argued that the $r$ round trips can be
``unfolded'' into one round trip in a system of size $rL$, hence a
proper Floquet ansatz should additionally include fractional momentum
states. We also showed that such fractional momentum states naturally
emerge as leading order corrections to the integer momentum states for
the $r>1$ resonances. Taking these into consideration, the proper
ansatz is
\begin{gather}
  \label{ansatz-frac-k}
  |\psi(t)\rangle = \sum_k \sum_{\rho = 0}^{r-1} f_{k,\rho}|k + \frac{\rho}{r}\rangle e^{-i(rk + \rho)\Omega t}\ ,
\end{gather}
where $|k + \frac{\rho}{r}\rangle$ are the fractional momentum states
Eq.~\ref{frac-k}. Note that their average energies do \emph{not} fall
on the dispersion curve of the integer momentum states. Instead, one
has ($\kappa = k + \frac{\rho}{r}$)
\begin{gather}
  \langle \kappa | \op H_0 | \kappa\rangle = 2 \sum_{x=1}^{L-1} \langle \kappa | x \rangle\langle x+1 | \kappa\rangle = 2 \cos q_{\kappa} \left\{ 1 + \frac{1}{\lp} \left[ \frac{\sin(2q_{\kappa} - 2\kappa \pi)}{\sin(2q_{\kappa})} - 1 \right] \right\}\ , \\
  \langle \kappa | \kappa\rangle = 1 - \frac{\sin \kappa \pi}{\lp \sin q_{\kappa}}\cos(\kappa \pi - q_{\kappa})\ , \\
\end{gather}
hence the energy of $|\kappa\rangle$ is
\begin{gather}
  \langle E\rangle_{\kappa} = \frac{\langle \kappa | \op H_0 | \kappa\rangle}{\langle \kappa | \kappa\rangle} = E_{\kappa} + \mu_{\kappa} \ , \\
  \label{mu-kap}
  \mu_{\kappa} = \frac{1}{\lp} \left[ \cos(q_{\kappa} - 2\kappa\pi) - \cos q_{\kappa}) \right] + \mathcal{O}(\lp^{-2})\ ,
\end{gather}
where $E_{\kappa} = 2\cos q_{\kappa}$ is the the dispersion relation
of the integer momentum states, and $\mu_{\kappa}$ is the deviation
$\langle E\rangle_{\kappa} - E_{\kappa}$.

Close to resonance, one can expand the (integer-$k$)
dispersion relation as
\begin{gather}
  \varepsilon_k = \varepsilon_{*} + r(k-k_{*})\Omega + \frac{u}{T}(k-k_{*})^2 + \cdots\ .
\end{gather}
It is useful to simplify $\mu_{\kappa}$ by replacing, in
Eq.~\ref{mu-kap}, $q_{\kappa} \rightarrow q_{*}$ (where
$q_{*} = q_{k_{*}} = k_{*} \pi / \lp$ is the interpolated wave vector
at the resonance center $k_{*}$), and
$2\kappa \pi \rightarrow 2\pi \frac{\rho}{r}$, yielding
\begin{gather}
  \label{mu-rho}
  \mu_{\rho} = \frac{1}{\lp} \left[ \cos(q_{*} - 2\pi \frac{\rho}{r}) - \cos q_{*} \right]\ ,
\end{gather}
i.e., the deviation $\mu_{\rho}$ depends only on the fractional part
$\rho$.  Introduce a composite index
\begin{gather}
  \label{j-def}
  j = rk + \rho\ ,
\end{gather}
$j$ labels the integer momentum states in the unfolded system (length
$rL$).  Then invoking Eq.~\ref{fs} on Eq.~\ref{ansatz-frac-k} leads to
the following eigenvalue problem,
\begin{gather}
  \sum_{j\pr} \left[ \vartheta_j\delta_{jj\pr} + g_{j-j\pr}V_{jj\pr} \right] f_{j\pr} = (\theta - \theta_{*}) f_j\ ,
\end{gather}
where
\begin{gather}
  \label{r-params}
  \vartheta_j = \mu_{\rho} + u(\frac{j}{r} - k_{*})^2\quad , \quad V_{jj\pr} = \langle k + \frac{\rho}{r} | \op V^{(j\pr - j)} | k\pr + \frac{\rho\pr}{r}\rangle T\quad , \quad \theta_{*} = \varepsilon_{*} T - 2\pi r k_{*}
\end{gather}
The effective model is thus a 1D ``lattice'' with ``unit cell'' label
$k$ and ``sublattice'' label $\rho$. The ``onsite potential''
$\vartheta$ remains quadratic, but has an additional
sublattice-dependent ``chemical potential'' $\mu_{\rho}$.

Before analyzing the effective model, we first discuss why the
\emph{apparent} recurrence time of the wave packet solutions for
$r > 1$ is $T/r$.  This behavior can be understood from the form of
the ansatz. Note that $|\psi(t)\rangle$ in Eq.~\ref{ansatz-frac-k} can
be separated into ``sublattice'' contributions,
$|\psi(t)\rangle = \sum_{\rho}|\psi_{\rho}(t)\rangle$, where
$|\psi_{\rho}(t)\rangle = \sum_k f_{k,\rho}|k+\frac{\rho}{r}\rangle
e^{-ij(k,\rho) \Omega t}$. Since by construction,
$|\psi_{\rho}(T/r)\rangle = e^{-i2\pi \rho/r}|\psi_{\rho}(0)\rangle$,
each ``sublattice'' recur after a fraction of drive period
$\frac{T}{r}$, but with different phase shift. Thus even though
rigorously speaking the full state $|\psi(t)\rangle$ does not recur
after $T/r$ due to the phase shifts (the exact recurrence time is
$T$), its spatial pattern does approximately return after $T/r$.

We now analyze the effective model assuming the drive has the form
$\op V(t) = 2g \cos(\Omega t) |L\rangle\langle L|$, that is, a
modulation on the \emph{last} site at the fundamental frequency. The
reason to modulate the last (instead of the first) site is to simplify
the expression for the fractional momentum states, see discussion at
the end of the last section. Then the drive only couples $j$ to
$j\pm 1$. The effective model becomes
\begin{gather}
  \label{eff-1d-tb}
  \left[ \mu_{\rho} + \frac{u}{r^2} \right] (j - j_{*})^2 f_j^{(\rho)} + \tau(f_{j+1}^{(\rho + 1)} + f_{j-1}^{(\rho - 1)}) = \lambda f_j\quad , \quad
\end{gather}
where $j_{*} = k_* / r$, $\lambda = (\theta - \theta_{*})$, and we
have used the approximation that the ``hopping'' $\tau$ is
``site''-independent. Note that we have placed a superscript $\rho$ to
the coefficients $f_j$, where $\rho = j \mod r$ (Eq.~\ref{j-def}), and
the superscripts are understood as carrying an implicit $\mod r$
(i.e., $\rho \pm 1$ should be understood as $(\rho \pm 1) \mod r$,
etc.). Let us now Fourier transform the index $j$ into a continuous
conjugate variable $y$,
\begin{gather}
  f_j^{(\rho)} \equiv \int dy \check f^{(\rho)}(y) e^{-i(j - j_{*}) y}\ .
\end{gather}
Note that the transformation is performed \emph{as if} $\rho$ is
\emph{independent} of $j$. What this means is that if one were given
$r$ continuous functions $\check f^{(\rho)}(y)$,
$\rho = 0, 1,\cdots, r-1$, then only Fourier components with
$j \equiv \rho \mod k$ are relevant as solution to
Eq.~\ref{eff-1d-tb}. In terms of $\check f^{(\rho)}$,
Eq.~\ref{eff-1d-tb} becomes a coupled Mathieu's equation,
\begin{gather}
  \label{coupled-mat}
  \left[  \op M - \frac{u}{r^2} \partial_y^2 \right] \vec{\check f} = \lambda \vec{\check f}\ ,
\end{gather}
where
\begin{gather}
  \op M = \begin{pmatrix}
      \mu_0 & \tau e^{-iy} & & & & \tau e^{iy}\\
      \tau e^{iy} & \mu_1 & \tau e^{-iy} \\
      & \tau e^{iy} & \mu_2 & \ddots\\
      & & \ddots & \ddots & \ddots\\
      & & & \ddots & \mu_{r-2} & \tau e^{-iy}\\
      \tau e^{-iy} && & & \tau e^{iy} & \mu_{r-1}
    \end{pmatrix}\quad , \quad \vec{\check f} =     \begin{pmatrix}
      \check f^{(0)} \\
      \check f^{(1)} \\
      \vdots \\
      \vdots \\
      \vdots \\
      \check f^{(r-1)}
    \end{pmatrix}
\end{gather}
The general strategy is then to solve Eq.~\ref{coupled-mat} in the
diagonal basis of the matrix $\op M$.

Since a generic $\op M$ cannot be diagonalized analytically, we will
specialize to $r=2$. In this case, one has
\begin{gather}
  r = 2 \implies \op M =
  \begin{pmatrix}
    \mu_0 & 2\tau \cos(y) \\
    2\tau \cos(y) & \mu_1
  \end{pmatrix}\ .
\end{gather}
Denoting the diagonal bases of $\op M$ as $\check f^{\pm}(y)$, then
Eq.~\ref{coupled-mat} becomes
\begin{gather}
  \left[- \frac{u}{r^2} \partial_y^2 \pm \sqrt{\frac{\Delta \mu^2}{4} + 4 \tau^2 \cos^2 y}\right] \check f^{\pm}(y) = (\lambda^{\pm} - \bar \mu)f^{\pm}(y)\quad , \quad \Delta \mu = \mu_1 - \mu_0\quad , \quad \bar \mu = \frac{\mu_0 + \mu_1}{2}\ .
\end{gather}
The problem is equivalent to a particle moving in a periodic potential
$U^{\pm}(y) = \pm \sqrt{\frac{\Delta \mu^2}{4} + 4\tau^2\cos^2y}$. The
Floquet wave packets correspond to bound states in one of the two
potentials. Near the bottom of either potential, one may Taylor expand
in $y$ and obtain
\begin{gather}
  U^+(y) \simeq \frac{\Delta \mu}{2} + \frac{4\tau^2}{\Delta \mu}
  \delta y^2 + \cdots\quad , \quad U^-(y) \simeq - \sqrt{\frac{\Delta
      \mu^2}{4} + 4\tau^2} + \frac{2\tau^2}{\sqrt{\frac{\Delta \mu^2}{4} + 4\tau^2}} \delta y^2 + \cdots\ .
\end{gather}
We expect Floquet wave packet solutions to be low-lying states of the
effective lattice model Eq.~\ref{eff-1d-tb} (this is because at higher
quasienergies, the ``hopping'' cannot efficiently mix neighboring
``sites'', hence the solutions there are closer to single-momentum
states, which are spatially extended). This means at a weak drive
strength (and hence small $\tau$), we should choose $U^-$ of the
two potential branches, as it has a negative overall shift. The
effective model is thus a \emph{continuum} harmonic oscillator of ``Hamiltonian''
\begin{gather}
  \label{sho-cont}
  \wt H = - m^{-1} \partial_y^2 + q \delta y^2  - C\ ,
\end{gather}
where the ``mass'' $m$, the ``stiffness'' $q$, and the constant shift
$C$ are
\begin{gather}
  m^{-1} = \frac{u}{r^2}\quad , \quad q = \frac{2\tau^2}{C}\quad ,\quad C = \sqrt{\frac{\Delta \mu^2}{4} + 4\tau^2}\ .
\end{gather}

The parameters $u, \tau$, and $\Delta \mu$ can be estimated as
follows. Parametrizing
\begin{gather}
  \beta_r  = \frac{r\lp}{T}\quad , \quad \gamma_r = \frac{1}{\sqrt{1 - \beta_r^2}}\ ,
\end{gather}
then from $\partial_k \varepsilon_k|_{k = k_{*}} = r\Omega$ and
$\frac{u}{T} = \frac{1}{2}\partial_k^2 \varepsilon_k |_{k = k_{*}}$, we have
\begin{gather}
  u = \frac{r^2\pi^2}{\beta_r^2 \gamma_r T}\ .
\end{gather}
For $r = 2$, from Eq.~\ref{r-params}, we can estimate $\tau$ as
\begin{gather}
  \tau = V_{j_{*}, j_{*} + 1} = g T \langle k_{*} |
  L\rangle\langle L | k_{*} + \frac{1}{2}\rangle = \frac{4 g}{\gamma_r}\ .
\end{gather}
Finally, using Eq.~\ref{mu-rho}, we have
\begin{gather}
  \Delta \mu = \mu_1 - \mu_0 = \frac{2 T}{\lp} \cos q_{*} = \frac{4}{\beta_r\gamma_r}\ .
\end{gather}
The ``frequency'' of the emergent harmonic oscillator,
Eq.~\ref{sho-cont}, is then
\begin{gather}
  \varpi  = 2 \sqrt{m^{-1} q} = \frac{8\pi g}{\gamma_r \sqrt{r\lp \sqrt{1 + 16g^2 \beta_r^2}}}\ .
\end{gather}
Note that at weak drive, $\varpi \propto g$. As the drive
becomes stronger, $\varpi \propto \sqrt{g}$. This is different
from the $r = 1$ cases (with arbitrary $s$), where
$\varpi \propto \sqrt{g}$ even at weak drive, see
Eq.~\ref{efreq-d-s}.

\begin{figure}
  \centering
  \includegraphics[trim=180 130 220 90,clip,width=.5\textwidth]{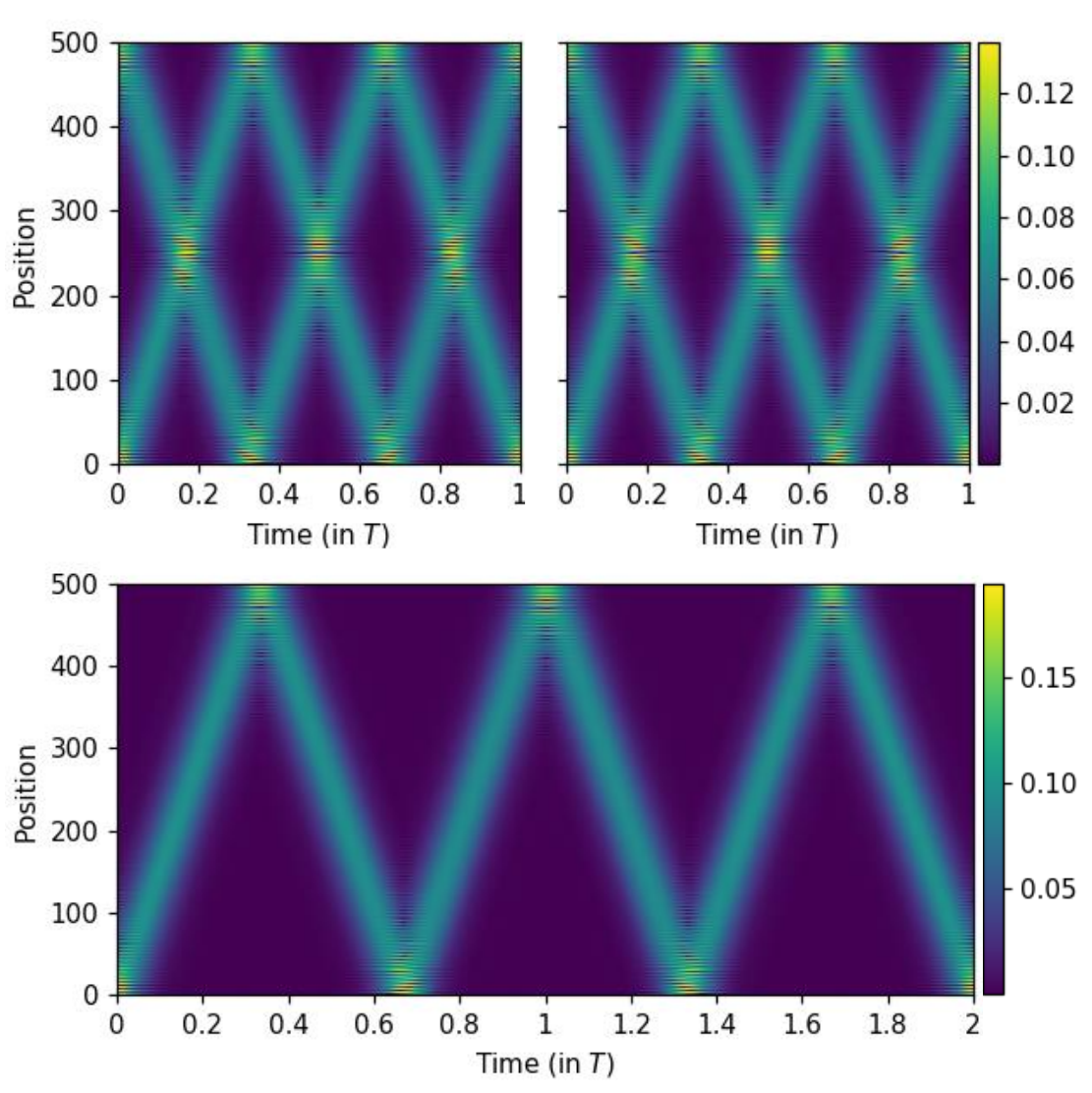}
  \caption{Non-dispersing wave packets with $s=2, r=3$ from
    Hamiltonian
    $\op H(t) = \sum_{x=1}^{L-1}|x\rangle\langle x+1| + h.c. +
    2g\cos(\Omega t)|1\rangle\langle 1|$, with system size $L = 500$,
    drive strength $g=1$, and drive period $T = 1005$
    ($\Omega = 2\pi/T$). Top: ``ground states'' of the two parity
    effective chains ($\sigma = 0, 1$). Their Floquet phases are
    $\pi + \delta$ apart, and numerically
    $\delta \simeq 1.62\times 10^{-5}\pi$. Bottom: Dynamical evolution
    of the time-crystalline recombination
    $|\phi^{0}\rangle = |\psi^{0}(0)\rangle +
    |\psi^{1}(0)\rangle$. After a tunneling time of
    $2\pi T/ \delta \simeq 1.23 \times 10^5 T$, it would evolve into
    the wave packet configuration of the opposite recombination
    $|\psi^{(0)}\rangle - |\psi^{(1)}\rangle$. }
  \label{fig:s2r3}
\end{figure}

\section{General $r \neq s \neq 1$ wave packets}
We briefly discuss the more general case of $r \neq s \neq 1$. In this
case, we can combine the two ansatze above and write
\begin{gather}
  |\psi^{\sigma}(t)\rangle = \sum_{\kappa,\rho} |k(\kappa, \rho,
  \sigma)\rangle e^{-i(r\kappa + \rho)\Omega t}  \ ,
\end{gather}
where $k(\kappa,\rho, \sigma)$ is a potentially fractional momentum,
\begin{gather}
  k(\kappa,\rho, \sigma) = s(\kappa + \frac{\rho}{r}) + \sigma  \ ,
\end{gather}
and $\kappa, \rho, \sigma$ are integers, with $\rho = 0, 1, \cdots, r$
and $\sigma = 0, 1, \cdots, \sigma$.  Thus invoking Eq.~\ref{fs} on
this ansatz will yield an effective lattice model of $s$ decoupled
chains (labeled by $\sigma$), each with $r$ sublattices (labeled by
$\rho$). Note that each chain (i.e., a specific $\sigma$) can be
analyzed in the same way as the $s = 1, r > 1$ case, except the index
$j$ in Eq.~\ref{j-def} is now $j = r\kappa + \rho$ (i.e., replace $k$
there by $\kappa$). Similar to the $r = 1$ case, the ``onsite''
energies of the $s$ chains have an equal spacing of $2\pi / s$ to
leading order (with higher order corrections arising from the coupling
between different $\sigma$ sectors), but otherwise essentially
identical, hence a given $(s,r)$ solution is necessarily one of $s$
partners with almost identical spatial-temporal patterns, and their
quasienergies are equally spaced by $\Delta\theta = 2\pi/s$ to leading
order. Combining the results of $s = 1$ and $r = 1$, one can see that
at $s \neq r \neq 1$, an $(s,r)$ Floquet eigenstate consists of $s$
wave packets, each completing a fraction $\frac{r}{s}$ of round trip
in one drive period. The individual wave packets can be resolved by
linear recombinations of the $s$ partners, similar to
Eq.~\ref{phi-resolv}, hence their true recurrence time is
$2\pi / \Delta \theta = sT$. However, since they completed $r$ round
trips in $sT$, their \emph{apparent} recurrence time is
$\Trec = sT/r$. In Fig.~\ref{fig:s2r3}, we plot the ``ground states''
of the two independent effective chains for $s = 2, r = 3$, and their
time-crystalline recombination. The latter completes $r=3$ round trips
in $s=2$ drive periods.

\subsection{Emergent lattice model with cubic potential}
\begin{figure}
  \centering
  \includegraphics[trim=160 130 200 90,clip,width=.5\textwidth]{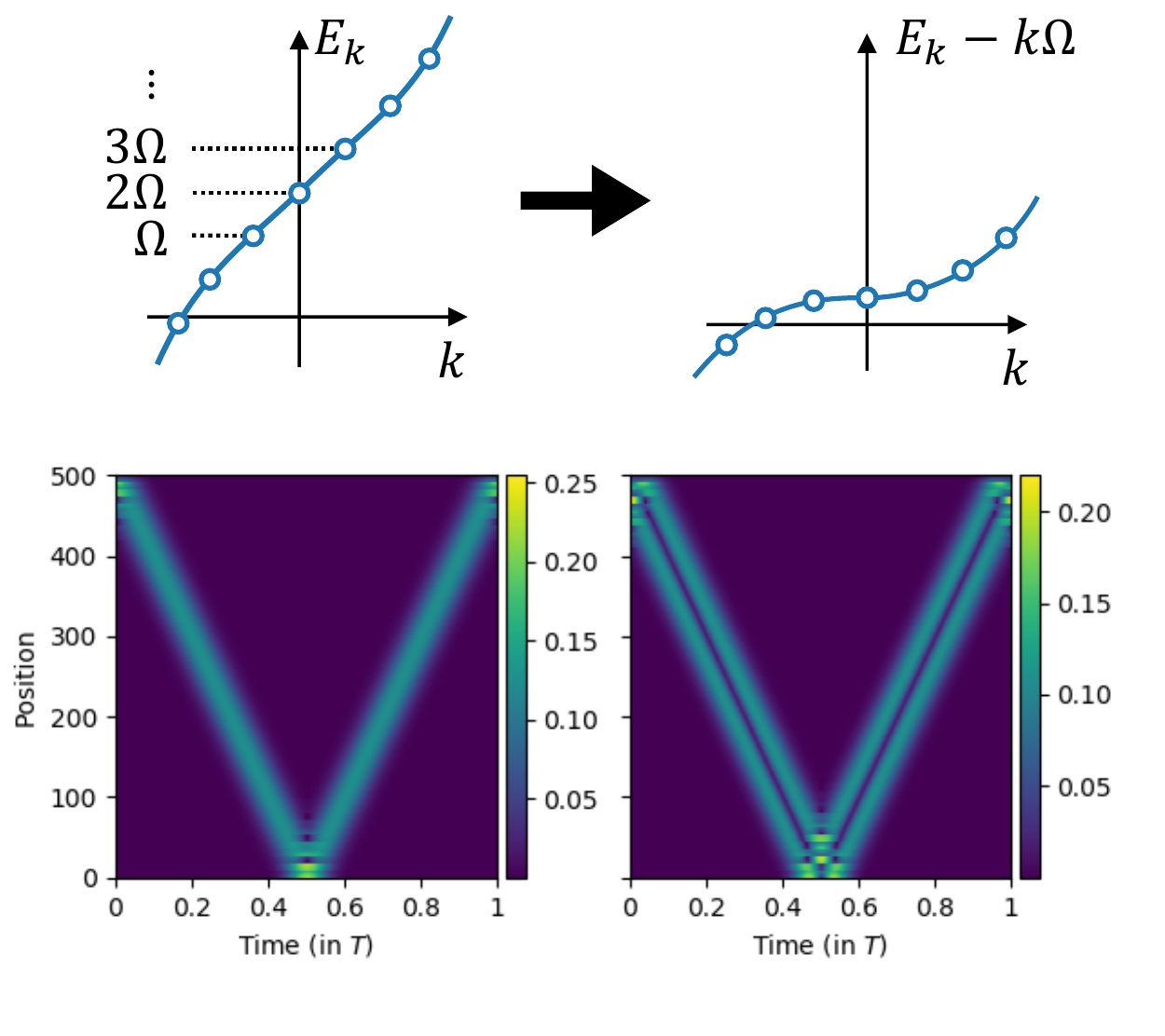}
  \caption{Non-dispersing wave packets from an effective ``cubic''
    oscillator. When the drive frequency resonates at the inflection
    point of a spectrum (top left), the effective model becomes a
    tight binding chain with a cubic onsite potential (top right).
    Bottom panels: Floquet wave packet solutions corresponding to
    eigenstates of the effective model closest to ``zero'' energy
    (i.e., center of the cubic potential). System size $L=500$, drive
    strength $g = 1$, drive period $T = 501$ (Setting $T=L+1$ matches
    the drive frequency exactly at the spectral inflection point).}
  \label{fig:cubic}
\end{figure}

Emergence of non-dispersing Floquet wave packets does not rely on the
``on-site potential'' in the effective lattice model being
quadratic. In this section, we discuss the case where the effective
potential becomes cubic. Such a scenario would arise, for example, by
fine-tuning the drive frequency to match the level spacing at the
inflection point of the undriven spectrum,
$T \simeq \wt L\Rightarrow q_{*} \simeq \pi$. See
Fig.~\ref{fig:cubic}.  By definition, the quadratic term in the Taylor
expansion of $\varepsilon_k$ vanishes, and one instead has
$\varepsilon_k = \varepsilon_{*} + \Omega (k-k_{*}) +
\frac{u}{T}(k-k_*)^3 + \cdots$.  The effective lattice model is now
$u(k-k_{*})^3 f_k + \tau(f_{k-1} + f_{k+1}) = (\theta - \theta_{*})
f_k$, where
$u = \frac{1}{6} T \frac{\partial^3 \varepsilon_k}{\partial
  k^3}|_{k_{*}} = \pi^3 / 3\tilde L^2$, while $\tau$ has the same
expression as in quadratic case and is $\tau = 2g$. %
Such an arrangement can host wave packet solutions, because as long as
$\tau$ is not too small, it can still efficiently couple several
nearby $k$ ``sites'' together. The potential profile only matters in
determining how many $k$ points can be coupled, and the weight
distribution among them. Note that while the quantum mechanical
problem of a particle in continuous space, with a purely cubic
potential, has no real eigenvalues \cite{ferreira14}, the discrete
nature of our effective model here places a natural cutoff on the
cubic potential (a ``site'' with too high a potential cannot couple to
neighboring sites via hopping)---in other words, the potential is
cubic near the center, but has effective infinite walls on both sides,
hence there is no subtlety in obtaining wave packet solutions with
real eigenvalues.  Indeed, similar to the quadratic case, the number
of wave packet states $D$ can be estimated as
$u|\delta k|^3 \le \tau \Rightarrow |\delta k| \le (\tau/u)^{1/3}
\Rightarrow D \simeq 2(\tau/u)^{1/3} \propto (g\tilde
L^2)^{1/3}$. Compared with the quadratic case, these wave packets have
a broader weight distribution in $k$ due to the flatter cubic
potential, leading to more compact coordinate space Floquet wave
packets.  The crossover drive strength is obtained by having $D = 2$
(instead of $1$, because the cubic model has a particle-hole
symmetry), and is thus $\tau_c \simeq u $, or
$g_c \propto \tilde L^{-2}$. To estimate the Floquet level spacing
near $\theta_{*}$ (the analogue of $\varpi$ in the quadratic case), we
use the effective Hamiltonian of a generic power law potential to
write $\Delta \theta^{(\nu)} = u \Delta X^\nu + \tau \Delta K^2$,
where $X$ and $K$ are the ``position'' and ``translation generator''
of the emergent lattice, and $\Delta X, \Delta K$ their
variances. Minimizing $\Delta \theta^{(\nu)}$ under the constraint of
minimal uncertainty $\Delta X \Delta K = 1$ ($\hbar = 1$) then leads
to the level spacing
\begin{gather}
  \varpi^{(\nu)} = \frac{\nu+2}{\nu}\tau \left[\frac{2\tau}{\nu u}\right]^{-\frac{2}{\nu+2}}\ .
\end{gather}
One can verify that $\varpi^{(2)}$ recovers the quadratic emergent
``frequency'' $\varpi$. For the cubic case, we have
$\varpi^{(3)} = \frac{5}{3} \left[ \frac{3}{2} \right]^{2/5} u^{2/5}
\tau^{3/5} \propto \left( g^3/\tilde L^4 \right)^{1/5}$. Since a tight
binding model with cubic potential has particle hole symmetry, its
eigenvalues come in $\pm$ pairs, and the analogue of ``low lying''
state are those with eigenvalues close to zero (i.e., potential
center). The bottom two panels in Fig.~\ref{fig:cubic} plots the two
lowest lying Floquet eigenstates (in the positive eigenvalue branch of
the effective model).

\subsection{Change of the width of a Floquet wave packet in one drive period}

In this section, we estimate how much the width of the (``ground
state'') Floquet wave packet (i.e., the one with a single spatial
peak) changes within one drive period. As shown in the main text, the
Floquet wave packets remain spatially compact at all time. Its typical
width can be estimated as
\begin{gather}
  W \simeq \frac{L}{D}\ ,
\end{gather}
where $L$ is chain length, and $D$ is the number of wave packet
states, which we have estimated in the main text. This is because
crudely speaking, one can think of all Floquet wave packet states as
spanning the same spatial extent, hence on average, each wave packet
has a spatial extent of $W$, which for the ``ground state'' is its
width.

To estimate how much the width changes, recall that $D$ was estimated
by finding how many momentum states can be efficiently coupled by the
Floquet drive --- the Floquet wave packets can be viewed as resulting
from a ``degenerate perturbation'' theory in the Hilbert space spanned
by these momentum states. Since we only apply drive on a single site,
the wave packet propagates mostly freely when away from the drive
site, hence its spread is due to the difference in velocity between
the fastest and slowest momentum component. To make the discussion
more general, we consider the following expansion of the (undriven)
energy spectrum,
\begin{gather}
  \varepsilon_k = \varepsilon_{*} + \Omega(k-k_{*}) + \frac{u_n}{T}(k-k_{*})^n + \cdots\ ,
\end{gather}
where $n = 2,3$ give the quadratic and cubic cases discussed
before. The effective lattice model is (see discussion in the main text and the previous section on the cubic case)
\begin{gather}
  u_n(k-k_{*})^n f_k + \tau (f_{k-1} + f_{k+1}) = (\theta - \theta_{*}) f_{k}\ .
\end{gather}
The number of wave packet states $D$ is estimated as the number of
(integer) $k$ ``sites'' that can be reached from $k_{*}$ by one
``hop'' $\tau$,
\begin{gather}
  u_n (\Delta k)^n = \tau\implies \Delta k = \left[ \frac{\tau}{u_n}  \right]^{1/n}\quad , \quad D = 2 \Delta k\ .
\end{gather}
This yields
\begin{gather}
  W \simeq \frac L D \simeq L \left[ \frac{u_n}{\tau}  \right]^{1/n}.
\end{gather}
The wave packet width is inversely proportional to the drive strength $\tau$, as expected. For the simple tight binding model we consider in the main text, $u_n$ scales as $1/L^{n-1}$, and hence $W\sim (L/\tau)^{1/n}$. This translates into  tighter wave packets for larger $n$.

The velocity of momentum component $k$ is
\begin{gather}
  v_k = \frac{L}{\pi} \frac{\partial \varepsilon_k}{\partial k} = \frac{2L}{T} + \frac{n u_n L}{\pi T}(k-k_{*})^{n-1}\ .
\end{gather}
The velocity difference between the fastest and slowest momentum component is then
\begin{gather}
  \Delta v = \frac{2 n u_n L}{\pi T} (\Delta k)^{n-1}\ .
\end{gather}
It takes the wave packet $T/2$ to traverse the chain from end to
end (assume $r = s = 1$ for this estimate), during which the slowest component will lag the fastest one by a
distance of
\begin{gather}
  \Delta W = \Delta v \, T/2 = \frac{n u_n L}{\pi}(\Delta k)^{n-1}\ ,
\end{gather}
which can be used to approximate the amount of change in the wave
packet width during one drive period. Physically, the wave packet
expands ``freely'' by $\Delta W$ when moving \emph{toward} the drive
site (at one end of the chain), and contracts when reflecting off the
driven site. Without the drive, the wave packet would continue to
expand after the reflection; the effect of the drive is to manipulate
the phase shifts in such a way as to reverse the interference effect of
reflection at the boundary.

It is interesting to note that the relative change in the wave packet
width, $\Delta W/W$, does \emph{not} depend on $u_n$, i.e., the
nonlinearity in the dispersion $\varepsilon_k$,
\begin{gather}
  \frac{\Delta W}{W} =  \Delta W \frac{D}{L} = \frac{2 n \tau}{\pi}\ .
\end{gather}
Recall that for $n = 2,3$, the effective ``hop'' $\tau$ can be
estimated as $ \tau = 2g (L+1)/T$ (and for $n=3$, i.e., for the
quadratic subleading term to vanish, one needs $L+1 = T$), thus the
relative spread scales as
$\Delta W/W \propto g v_g$, where $v_g = 2(L+1)/T$ is the 
group velocity of the wave packet.

Note that in the undriven Hamiltonian $\op H_0 = h\sum_x |x\rangle \langle x+1| + h.c. $, we have assumed the hopping $h = 1$ when deriving the above results. If a generic, dimensionful $h$ is reinstated, then in the expression for the dimensionless $\tau$, one would find $g$ and $T$ to be replaced by their dimensionless versions, $\wt g = g/h$ and $\wt T = hT$. The width ratio becomes $\Delta W / W = \frac{4n}{\pi}\frac{\wt g(L+1)}{\wt T}\propto g v_g/h^2$, which is dimensionless ($g, h$, and $v_g$ all having the dimension of inverse time, $L$ is a dimensionless integer).

\end{document}